\documentclass[journal=aamick,manuscript=article]{achemso}

\usepackage[version=3]{mhchem} 
\usepackage{hyperref}

\setkeys{acs}{maxauthors=999}
\setkeys{acs}{etalmode=truncate}

\usepackage[
    type={CC},
    modifier={by},
    version={3.0},
]{doclicense}

\title{Domain wall migration-mediated ferroelectric switching and Rashba effect tuning in GeTe thin films}


\author{Libor Vojáček}
\affiliation{Univ. Grenoble Alpes, CEA, CNRS, SPINTEC, 38054 Grenoble, France}

\author{Mairbek Chshiev}
\affiliation{Univ. Grenoble Alpes, CEA, CNRS, SPINTEC, 38054 Grenoble, France}
\altaffiliation{Institut Universitaire de France, Paris, 75231, France}

\author{Jing~Li}
\email{Jing.Li@cea.fr}
\affiliation{Université Grenoble Alpes, CEA, Leti, F-38000, Grenoble, France}
\altaffiliation{European Theoretical Spectroscopy Facility (ETSF), F-38000 Grenoble}

\begin{document}

\begin{tocentry}
\includegraphics[width=0.8\columnwidth]{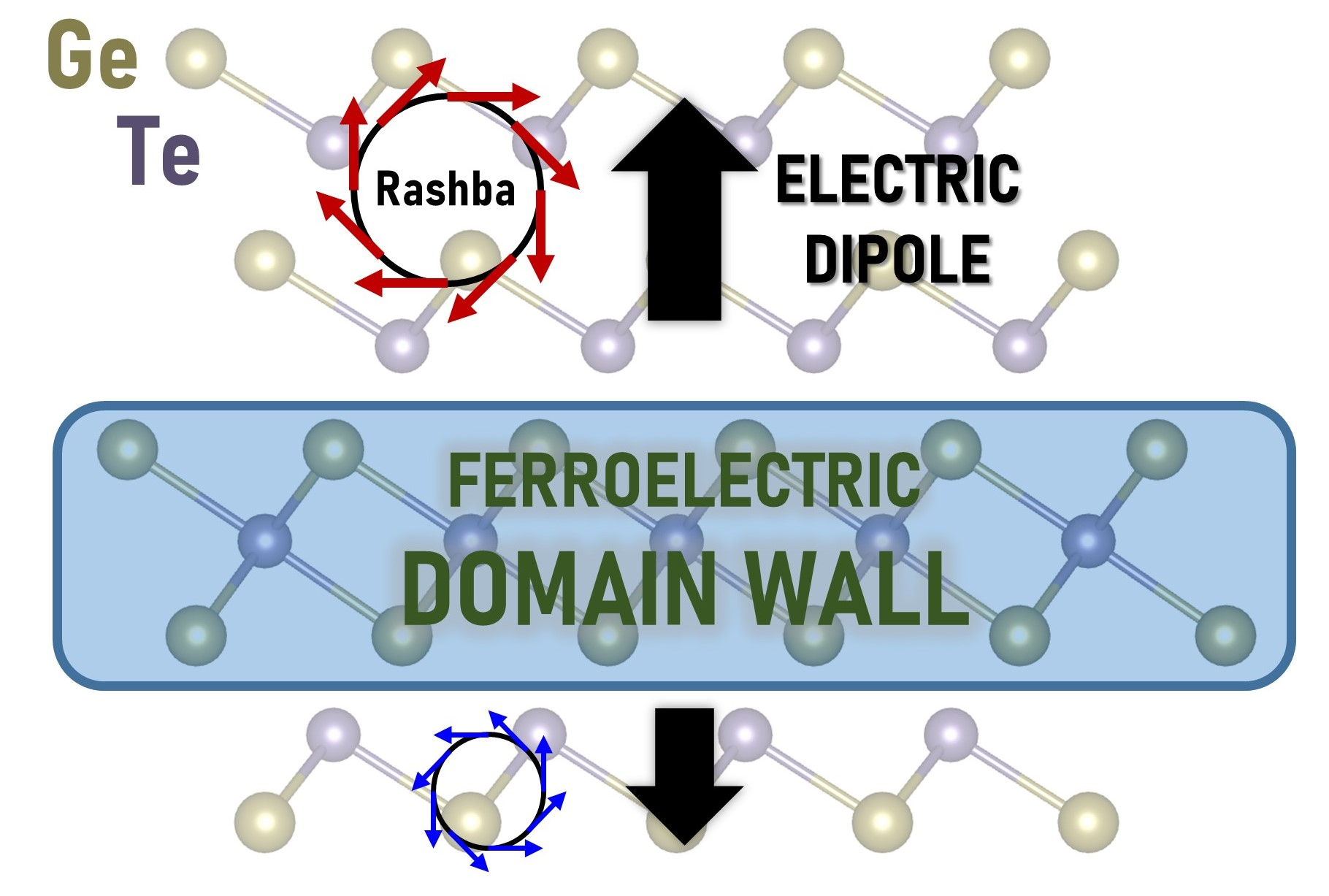}
\end{tocentry}

\begin{abstract}
Germanium Telluride (GeTe), identified as a ferroelectric Rashba semiconductor, is a promising candidate for future electronic devices in computing and memory applications. 
However, its ferroelectric switching on a microscopic scale remains to be understood.
Here, we propose that the migration of a domain wall can be the mechanism that mediates the ferroelectric switching.
By employing $ab~initio$ methods, such a mechanism is characterized by an energy barrier of $66.8$ meV/nm$^2$, in a suitable range for retention and switchability.
In accompanying the domain wall migration, the net Rashba effect is tunable, as it is a result of competition between layers with opposite electric polarization. 
These results shed light on the ferroelectric switching mechanism in GeTe, paving stones for the design of potential GeTe-based devices.  
\end{abstract}

\section{Introduction}
The recent discovery of the ferroelectric Rashba semiconductor \cite{DiSante_2013}, represented by a prototypical material GeTe \cite{Pawley_1966}, opens a novel path in the design of computing and memory devices \cite{Rinaldi_2016,Rinaldi_2018,Wang_2020,Varotto_2021}.
The ferroelectricity in GeTe, tunable by an external electric field, demonstrated non-volatility and remarkable retention at room temperature.\cite{Rinaldi_2018}
More importantly, with the strong coupling to the Rashba effect, ferroelectricity effectively steers and manipulates spin transport, thanks to innated spin-momentum locking in the Rashba effect.
Benefiting from such a mechanism, spin-charge interconversion, which connects conventional electronic devices and up-to-date spintronic devices, is demonstrated in GeTe at room temperature \cite{Rinaldi_2016, Varotto_2021}.

Ferroelectric switching is a rearrangement of atoms, which turns over the electric dipole. 
The energy barrier during the switching determines the data retention and writing speed, which are crucial for memory applications. 
However, the energy barrier in GeTe ferroelectric switching is not well investigated, and the switching mechanism in the microscopy scale remains to be explored.
The ferroelectric switching in bulk $\alpha$-GeTe (rhombohedral phase) might be due to the motion of the atomic plane (Ge and Te), which is controlled by an external electric field \cite{Kolobov_2014}.
In GeTe thin film, the atoms on the top and bottom surface are responsible for contact with other materials, such as electrodes, and are unlikely to participate in ferroelectric switching \cite{Meng_2017}.
Phase-charge in GeTe film between rhombohedral and monoclinic is proposed \cite{Jeong_2021}.
The other possible mechanism could be from the domain walls \cite{Nukala_2017, Croes_2021}.

In the following, we propose a GeTe thin film model that demonstrates the ferroelectric switching facilitated by the domain wall migration, which is controllable by an external electric field. 
The energy barrier of the switching mechanism is extracted using the $ab~initio$ method and compared with the ferroelectric switching in bulk $\alpha$-GeTe and GeTe monolayer. 
Furthermore, we investigated the Rashba effect during the domain wall migration to verify its tunability. 

\section{Method}

Density functional theory (DFT) calculations are performed using the Vienna Ab initio Simulation Package (VASP) \cite{Kresse_1993, Kresse_1996a, Kresse_1996b, Kresse_1999} with Perdew–Burke-Ernzerhof (PBE) functional \cite{Perdew_1996} and Grimme's D3 dispersion correction \cite{Grimme_2010}.
The Brillouin zone of GeTe thin film is sampled by the $7\times 7 \times 1 $ Gamma-centered $k$-mesh.
A $1.5$ nm thick Vacuum was inserted in film calculations.
A plane-wave basis set is used with a cut-off energy of $400$ eV.
The convergence for electronic density is set to $10^{-8}$ eV.
Spin-orbit coupling is taken into account in the band structure calculation.
The atomic structure is relaxed with the threshold energy of $10^{-5}$ eV. 
The migration of the domain wall is computed using the climbing-image nudged elastic band (NEB) method \cite{Henkelman_2000,Sheppard_2012} implemented in Transition State Tools for VASP.

\section{Results and Discussion}

\begin{figure*}[t]
\includegraphics[width=1\columnwidth]{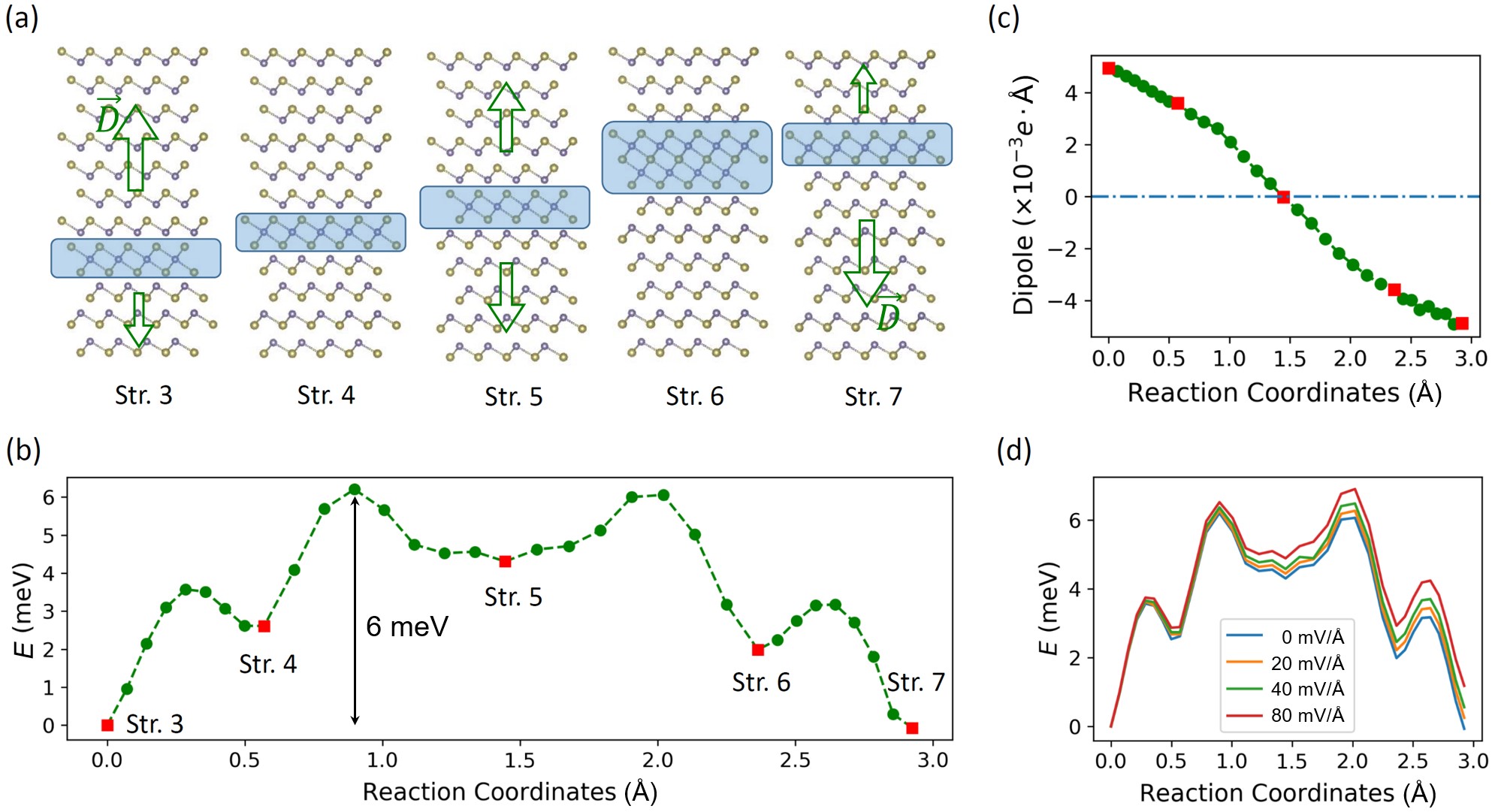}
\caption{Ferroelectric switching by the migration of domain-wall demonstrated in a $4$ nm thick Te-riched GeTe film (Ge$_{11}$Te$_{12}$).
  (a) Atomistic structure with domain wall (shaded with light blue) in various positions of the thin film. Electric polarization is the opposite for layers above and below the domain wall. These structures are labeled from 3 to 7, which indicates the number of monolayers below the domain wall. 
  (b) The energy along the domain-wall migration path from Str.~3 to Str.~7. Three local minima are at Strs. 4, 5 and 6. The global energy barrier is about $6$ meV per unit cell, around $66.8$ meV/nm$^2$.
  (c) The electric dipole along the domain-wall migration. Structure 5, with the domain wall at the film's center, has zero electric dipole.
  (d) Energy measured with respect to Str. 3 along the domain wall migration path under electric fields.  
}
\label{fig:dwall}
\end{figure*}

The Te-terminated surface is $60$ meV/\AA$^2$ more stable than the Ge-terminated one in the $\alpha$-GeTe thin film \cite{Deringer_2012}, which might be the reason for p-type doping observed in experiments \cite{Krempaský_2016,Yang_2021,Varotto_2021}.
Therefore, the initial structure of the GeTe thin film is constructed from the $\alpha$-GeTe crystal structure with the $z$ axis along the $[0001]$ direction terminated with Te atom for both top and bottom surfaces.
The Te-riched GeTe thin film is about $4$ nm thick and contains 12 Te and 11 Ge atoms in the unit cell with an in-plane lattice parameter $4.24$~\AA and an angle of $60$ degrees (unit cell area of $8.98$~\AA$^2$).
By relaxing the initial structure, a domain wall is noticed in the thin film, as shown by Str.~3 in Fig.~\ref{fig:dwall}(a), which is similar to the structure presented in Ref.~\cite{Deringer_2012}.
The domain wall has a cubic structure with an identical Ge-Te bond length on the top and bottom sides, separating the film into top and bottom layers with electric dipoles pointing up and downward, respectively.
By adjusting the inter-layer distance between Ge and Te atomic planes in the initial structure and relaxing it, domain walls at various positions in the film are obtained (Fig.~\ref{fig:dwall}(a)).
Among these structural relaxations, even though trying to insert the domain wall above the first GeTe monolayer, we noticed that the domain wall tends to keep a distance from the surface, i.e., three GeTe monolayers. 
Additionally, the domain wall could be triple or quintuple atomic layers using a maximum length of $3$~\AA~for Ge-Te bonds.

Figure~\ref{fig:dwall}(b) shows the relative energy along the domain wall migration path obtained by the NEB method using the structures relaxed in the previous step as initial and final configurations and seven images between them. 
The energy of the previously relaxed structures (indicated by red markers) are local minimums in the migration path.
However, we noticed that Str.~4 and Str.~6 are not symmetric. This can be the result of several local minima.
Structure 6, with a domain wall of quintuple atomic layer, has slightly lower energy than Str.~4.
The global migration path from Str.~3 to Str.~7 consists of four successive local migrations. Each local migration corresponds to the domain wall moving over one GeTe monolayer.  
The energy barrier is about $4$ meV per unit cell (UC) for the first two local migrations (Str.~3 to Str.~4, and Str.~4 to Str.~5).
The energy barrier decreases to about $2$ meV/UC for the last two local migrations (Str.~3 to Str.~4, and Str.~4 to Str.~5) because the initial local configuration has energy about $2$ meV/UC higher than the final local configuration. 
The global energy barrier is $6$ meV/UC (or $66.8$ meV/nm$^2$) from Str.~3 to Str.~7 in this $4$ nm thick GeTe film.
Such an energy barrier is significantly lower than ferroelectric switching in bulk and monolayer GeTe, which is discussed later in detail.

During the domain wall migration from Str. 3 to Str. 7, the bottom layer expands while the top one shrinks. Such a fact is manifested by the electric dipole $D$ (along $z$ direction) shown by Fig.~\ref{fig:dwall}(c), which decreases along the migration path. 
At Str. 5, the electric dipole crosses zero due to the cancelation of the dipole from the top and bottom layers, as they are the same size.
Under a vertical electric field $F_{\perp}$, the additional energy along the migration path is estimated as $\Delta E = D \cdot F_{\perp}$ under linear response, which is confirmed by the DFT simulations under electric field.
The energy along the migration path is therefore tilted, as shown in Fig.~\ref{fig:dwall}(d), implying that ferroelectric switching in GeTe thin film is controllable by an external electric field.
In experiments, the domain wall-mediated ferroelectric switching may end with an intermediate state with local minima and incomplete switching. The experimental technique, such as transmission electron spectroscopy, can identify the domain wall directly \cite{Gao_2011}. 

\begin{figure}
\includegraphics[width=0.8\columnwidth]{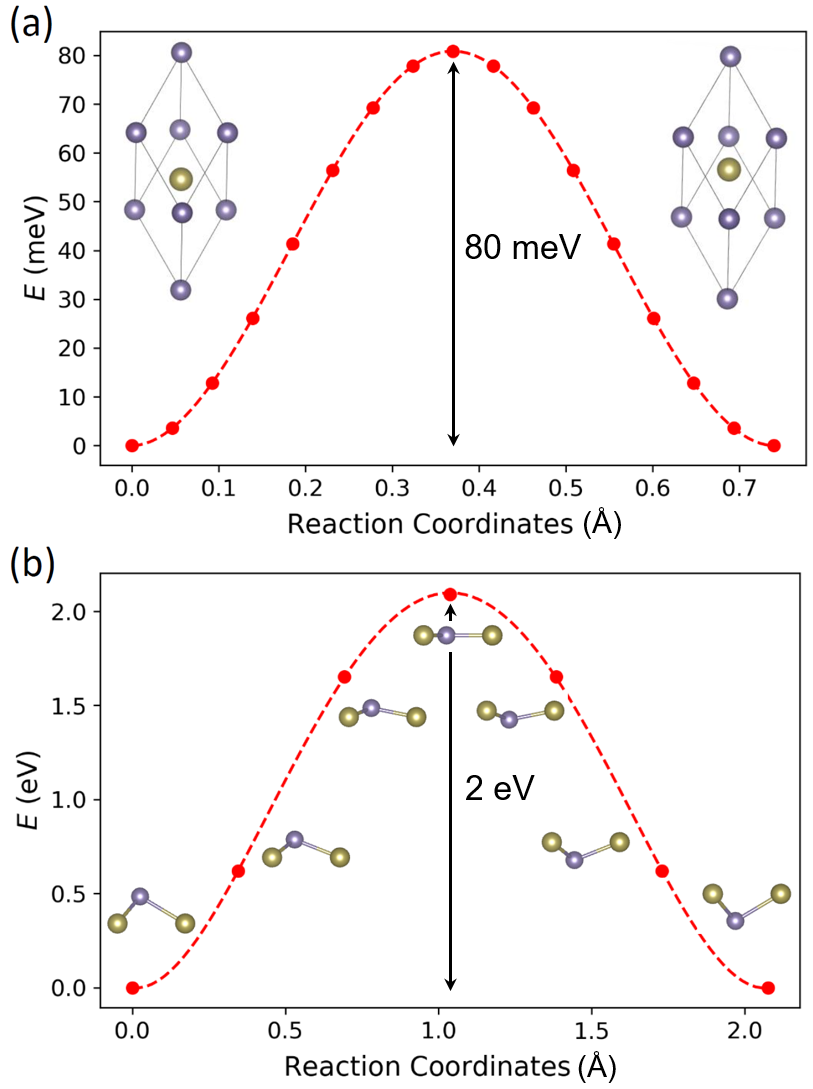}
\caption{
  The energy along the reaction coordinates during the ferroelectric switching in (a) bulk $\alpha$-GeTe and (b) GeTe monolayer.
  Inserts in (a) shows the unit cell at the initial and final states.
  Inserts in (b) are the unit cells of the GeTe monolayer, corresponding to the red markers.  
}
\label{fig:barrier}
\end{figure}

For comparison, we evaluate the energy barrier of ferroelectric switching in bulk $\alpha$-GeTe and GeTe monolayer.
In the relaxed structure of bulk $\alpha$-GeTe, the Te atom is located slightly off the center of the UC. Such asymmetry gives rise to the electric dipole. The ferroelectric switching is the motion of the Te atom to the other equilibrium position, giving the opposite dipole, as shown by the insert of Fig.~\ref{fig:barrier}(a). Such movement encounters an energy barrier of $80$ meV/UC (Fig.~\ref{fig:barrier}(a)), i.e., 1.4 eV/nm$^3$, which is about $5.6$ eV/nm$^2$ if we consider a $4$ nm thick GeTe film.
In the GeTe monolayer, the ferroelectric switching requires the interchange of the Ge and Te atomic planes, resulting in an inevitable cross-over of the two atomic planes, which gives the energy barrier of $2$ eV/UC, i.e. $22.26$ eV/nm$^2$. 
With such a high energy barrier, the GeTe monolayer is hardly switchable.
In contrast, the proposed ferroelectric switching by domain wall migration with barrier $66.8$ meV/nm$^2$ is much lower than the bulk $\alpha$-GeTe and GeTe monolayers.
With a surface area of $4$ nm$^2$, the barrier energy is one order of magnitude larger than thermal energy at room temperature, guaranteeing the retention of the ferroelectric state.
In addition, the ferroelectric switching remains controllable by an external field, as it does not depend on the surface area. 

From experiments up to now, the ferroelectric switching in GeTe thin film is likely to occur in two steps. First, ferroelectric switching occurs in the vertical direction between the top and bottom electrodes, forming a column-like domain. This could be due to the inhomogeneous electric field in the sample; some parts are easier to switch than others. The switched column-like domain contacts laterally with the remaining unswitched material. In the second step, the domain wall propagates laterally. The proposed domain wall migration-mediated ferroelectric switching is focused on the first step. However, more experimental evidence is required to verify the proposed mechanism by checking 1) if incomplete ferroelectric switching is allowed by using a slow pulse? 2) does Ge-Te to substrate chemical bonding participate in ferroelectric switching? 3) the ferroelectric switching barrier experimentally. 

\begin{figure*}
\includegraphics[width=1\columnwidth]{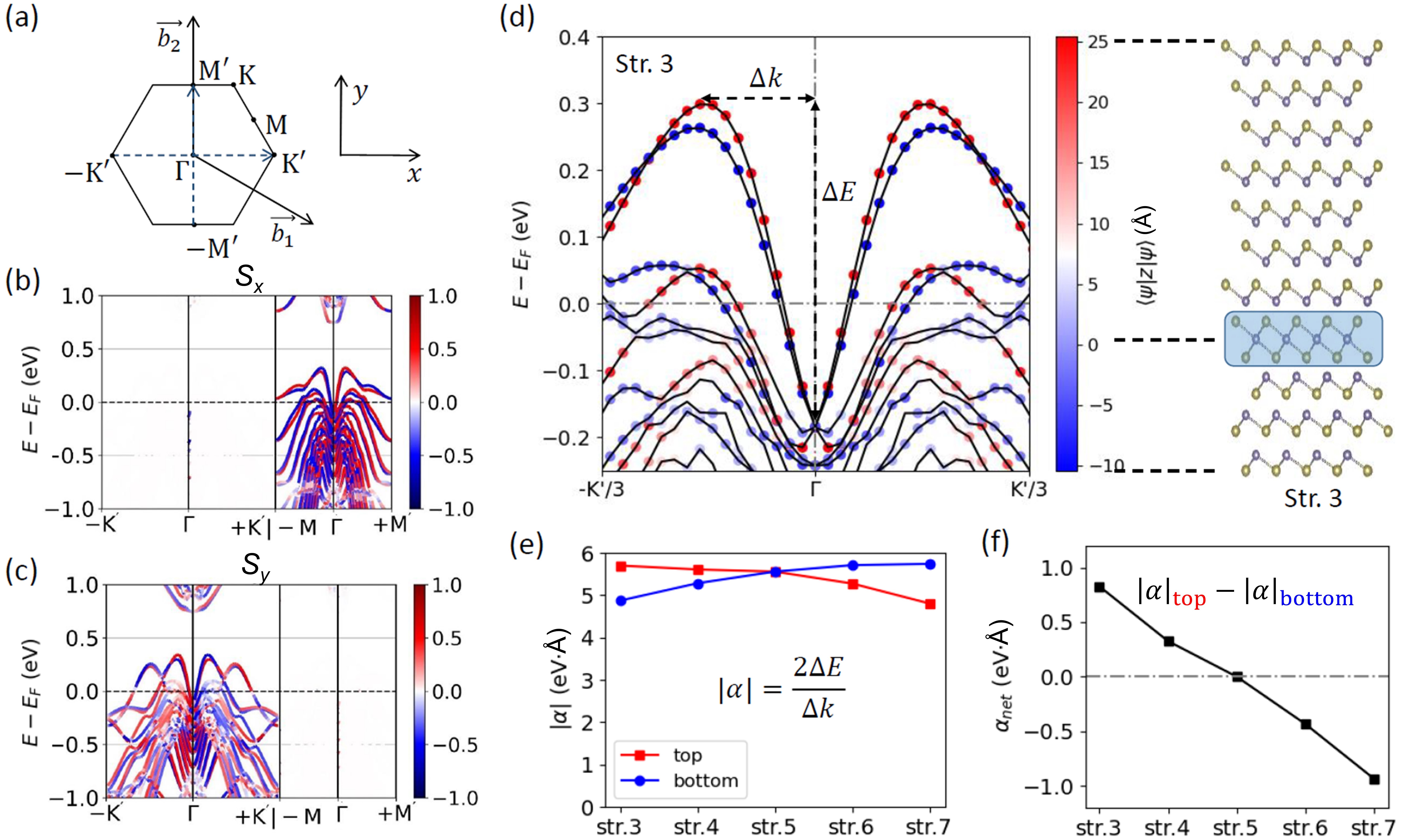}
\caption{Electronic structure and Rashba coefficient in Te-riched GeTe thin films. 
  (a) Brillouin Zone, the -K' to K' is along the $x$-axis, while -M' to M' is along the $y$-axis. 
  Electronic structure with spin quantization axis (b) along the $x$-axis and (c) along the $y$-axis. (b) and (c) shows the spin is perpendicular to the momentum, i.e., spin-momentum locking.
  (d) Zoomed electronic structure around $\Gamma$ and Fermi-level. Color markers indicated the expected position in the $z$-axis with the reference position at the domain wall center. 
  (e) Rashba coefficients for top and bottom layers extracted from electronic structure and (f) The net Rashba coefficients (the difference between top and bottom layers) versus structures along the domain wall migration.}
\label{fig:spin}
\end{figure*}

The electronic band structure is investigated to reveal the Rashba effect during the domain wall migration.
Taking Str. 3 as an example, the spin-momentum locking in the Rashba effect is demonstrated by the electronic structure plot with spin quantization axis along $x$ ($S_x$), and along $y$ ($S_y$). The spin component is perpendicular to the momentum, i.e. $k$ along $y$ ($\Gamma$ to $M'$) for $S_x$ (Fig.~\ref{fig:spin}(b)), and $k$ along $x$ for $S_y$ (Fig.~\ref{fig:spin}(c)).

In stoichiometry preserved GeTe thin film, namely the absence of the domain wall, the top of the valence band demonstrates only a single Rashba M-shape band \cite{Rinaldi_2018}.
However, two quasi-degenerate Rashba bands are identified at the valence band top in our model.  
To understand the nature of these two bands, we determine the expected position of wavefunctions in the $z$ direction, i.e., $\langle \psi|z|\psi \rangle$.
As shown by Fig.~\ref{fig:spin}(d), these two bands originated from the top and bottom layers, respectively.
From the band structure, the Rashba coefficient is evaluated \cite{Picozzi_2014}:
\begin{equation}
    |\alpha| = \frac{2\Delta E}{\Delta k},
\end{equation}
where $\Delta k$ and $\Delta E$ are the momentum and the energy measured from the top of a band to $\Gamma$, as denoted in Fig.~\ref{fig:spin}(d).
The top layer in Str.~3 has a more significant Rashba effect than the bottom layer as it is thicker. They become equal when the domain wall migrates to the center of the film (Str.~5), as shown in Fig.~\ref{fig:spin}(e).
The absolute value of Rashba coefficients is comparable to the previously reported values from theory and experiments \cite{Picozzi_2014, Liebmann_2016,Krempaský_2016,Yang_2021}.
However, because of the reserved chirality (shown in Fig.~\ref{fig:spin}(b) and (c)), the Rashba coefficient has an opposite sign for the top and bottom layers.
Therefore, a cancelation of the spin current takes place, leading to a net Rashba effect, with the coefficient as the difference between the top and bottom layers (Fig.~\ref{fig:spin}(f)).
Even though the net Rashba effect is small, the band structure is preserved to a Rashba M-shape.
This may lead to difficulty in extracting Rashba coefficients experimentally.
For angle-resolved photoemission spectroscopy (ARPES), it requires resolving the quasi-degenerate bands, and the bottom layer is not easy to reach.
Meanwhile, transport measurement accesses only the net Rashba effect. 

\section{Conclusion}

The domain wall migration by pushing the boundary toward the top or bottom surface is a plausible ferroelectric switching mechanism in GeTe thin film, as it is favorable energetically and controllable by an external electric field.
Two quasi-degenerated Rashba bands are identified at the top of the valence band.
The two bands belong to the two parts of the GeTe layer (top and bottom), separated by the domain wall. 
These two parts have an opposite electric dipole and opposite Rashba effect.
The strength of the Rashba effect increases in a given domain when the domain expands (upon the domain wall motion).
Therefore, the net Rashba effect is tunable by controlling the ferroelectricity.

\begin{acknowledgement}
The authors thank L. Vila and J.P. Attané for fruitful discussions.
Calculations were performed on computational resources provided by GENCI–IDRIS (Grant 2023-A0130912036 and 2024-A0150912036). 
This project has received funding from the \textit{European Union's Horizon 2020 research and innovation programme} under grant agreement No \textit{800945} — NUMERICS — H2020-MSCA-COFUND-2017.
\end{acknowledgement}

\bibliography{GeTe_acs}

\providecommand{\latin}[1]{#1}
\makeatletter
\providecommand{\doi}
  {\begingroup\let\do\@makeother\dospecials
  \catcode`\{=1 \catcode`\}=2 \doi@aux}
\providecommand{\doi@aux}[1]{\endgroup\texttt{#1}}
\makeatother
\providecommand*\mcitethebibliography{\thebibliography}
\csname @ifundefined\endcsname{endmcitethebibliography}
  {\let\endmcitethebibliography\endthebibliography}{}
\begin{mcitethebibliography}{26}
\providecommand*\natexlab[1]{#1}
\providecommand*\mciteSetBstSublistMode[1]{}
\providecommand*\mciteSetBstMaxWidthForm[2]{}
\providecommand*\mciteBstWouldAddEndPuncttrue
  {\def\EndOfBibitem{\unskip.}}
\providecommand*\mciteBstWouldAddEndPunctfalse
  {\let\EndOfBibitem\relax}
\providecommand*\mciteSetBstMidEndSepPunct[3]{}
\providecommand*\mciteSetBstSublistLabelBeginEnd[3]{}
\providecommand*\EndOfBibitem{}
\mciteSetBstSublistMode{f}
\mciteSetBstMaxWidthForm{subitem}{(\alph{mcitesubitemcount})}
\mciteSetBstSublistLabelBeginEnd
  {\mcitemaxwidthsubitemform\space}
  {\relax}
  {\relax}

\bibitem[Di~Sante \latin{et~al.}(2013)Di~Sante, Barone, Bertacco, and
  Picozzi]{DiSante_2013}
Di~Sante,~D.; Barone,~P.; Bertacco,~R.; Picozzi,~S. Electric Control of the
  Giant Rashba Effect in Bulk GeTe. \emph{Advanced Materials} \textbf{2013},
  \emph{25}, 509–513\relax
\mciteBstWouldAddEndPuncttrue
\mciteSetBstMidEndSepPunct{\mcitedefaultmidpunct}
{\mcitedefaultendpunct}{\mcitedefaultseppunct}\relax
\EndOfBibitem
\bibitem[Pawley \latin{et~al.}(1966)Pawley, Cochran, Cowley, and
  Dolling]{Pawley_1966}
Pawley,~G.~S.; Cochran,~W.; Cowley,~R.~A.; Dolling,~G. Diatomic Ferroelectrics.
  \emph{Physical Review Letters} \textbf{1966}, \emph{17}, 753–755\relax
\mciteBstWouldAddEndPuncttrue
\mciteSetBstMidEndSepPunct{\mcitedefaultmidpunct}
{\mcitedefaultendpunct}{\mcitedefaultseppunct}\relax
\EndOfBibitem
\bibitem[Rinaldi \latin{et~al.}(2016)Rinaldi, Rojas-Sánchez, Wang, Fu,
  Oyarzun, Vila, Bertoli, Asa, Baldrati, Cantoni, George, Calarco, Fert, and
  Bertacco]{Rinaldi_2016}
Rinaldi,~C.; Rojas-Sánchez,~J.~C.; Wang,~R.~N.; Fu,~Y.; Oyarzun,~S.; Vila,~L.;
  Bertoli,~S.; Asa,~M.; Baldrati,~L.; Cantoni,~M.; George,~J.-M.; Calarco,~R.;
  Fert,~A.; Bertacco,~R. Evidence for spin to charge conversion in GeTe(111).
  \emph{APL Materials} \textbf{2016}, \emph{4}, 032501\relax
\mciteBstWouldAddEndPuncttrue
\mciteSetBstMidEndSepPunct{\mcitedefaultmidpunct}
{\mcitedefaultendpunct}{\mcitedefaultseppunct}\relax
\EndOfBibitem
\bibitem[Rinaldi \latin{et~al.}(2018)Rinaldi, Varotto, Asa, Sławińska, Fujii,
  Vinai, Cecchi, Di~Sante, Calarco, Vobornik, Panaccione, Picozzi, and
  Bertacco]{Rinaldi_2018}
Rinaldi,~C.; Varotto,~S.; Asa,~M.; Sławińska,~J.; Fujii,~J.; Vinai,~G.;
  Cecchi,~S.; Di~Sante,~D.; Calarco,~R.; Vobornik,~I.; Panaccione,~G.;
  Picozzi,~S.; Bertacco,~R. Ferroelectric Control of the Spin Texture in GeTe.
  \emph{Nano Letters} \textbf{2018}, \emph{18}, 2751–2758\relax
\mciteBstWouldAddEndPuncttrue
\mciteSetBstMidEndSepPunct{\mcitedefaultmidpunct}
{\mcitedefaultendpunct}{\mcitedefaultseppunct}\relax
\EndOfBibitem
\bibitem[Wang \latin{et~al.}(2020)Wang, Gopal, Picozzi, Curtarolo,
  Buongiorno~Nardelli, and Sławińska]{Wang_2020}
Wang,~H.; Gopal,~P.; Picozzi,~S.; Curtarolo,~S.; Buongiorno~Nardelli,~M.;
  Sławińska,~J. Spin Hall effect in prototype Rashba ferroelectrics GeTe and
  SnTe. \emph{npj Computational Materials} \textbf{2020}, \emph{6}, 1–7\relax
\mciteBstWouldAddEndPuncttrue
\mciteSetBstMidEndSepPunct{\mcitedefaultmidpunct}
{\mcitedefaultendpunct}{\mcitedefaultseppunct}\relax
\EndOfBibitem
\bibitem[Varotto \latin{et~al.}(2021)Varotto, Nessi, Cecchi, Sławińska,
  Noël, Petrò, Fagiani, Novati, Cantoni, Petti, Albisetti, Costa, Calarco,
  Buongiorno~Nardelli, Bibes, Picozzi, Attané, Vila, Bertacco, and
  Rinaldi]{Varotto_2021}
Varotto,~S.; Nessi,~L.; Cecchi,~S.; Sławińska,~J.; Noël,~P.; Petrò,~S.;
  Fagiani,~F.; Novati,~A.; Cantoni,~M.; Petti,~D.; Albisetti,~E.; Costa,~M.;
  Calarco,~R.; Buongiorno~Nardelli,~M.; Bibes,~M.; Picozzi,~S.; Attané,~J.-P.;
  Vila,~L.; Bertacco,~R.; Rinaldi,~C. Room-temperature ferroelectric switching
  of spin-to-charge conversion in germanium telluride. \emph{Nature
  Electronics} \textbf{2021}, \emph{4}, 740–747\relax
\mciteBstWouldAddEndPuncttrue
\mciteSetBstMidEndSepPunct{\mcitedefaultmidpunct}
{\mcitedefaultendpunct}{\mcitedefaultseppunct}\relax
\EndOfBibitem
\bibitem[Kolobov \latin{et~al.}(2014)Kolobov, Kim, Giussani, Fons, Tominaga,
  Calarco, and Gruverman]{Kolobov_2014}
Kolobov,~A.~V.; Kim,~D.~J.; Giussani,~A.; Fons,~P.; Tominaga,~J.; Calarco,~R.;
  Gruverman,~A. Ferroelectric switching in epitaxial GeTe films. \emph{APL
  Materials} \textbf{2014}, \emph{2}, 066101\relax
\mciteBstWouldAddEndPuncttrue
\mciteSetBstMidEndSepPunct{\mcitedefaultmidpunct}
{\mcitedefaultendpunct}{\mcitedefaultseppunct}\relax
\EndOfBibitem
\bibitem[Meng \latin{et~al.}(2017)Meng, Bai, Gao, Gong, Wang, Duan, and
  Chu]{Meng_2017}
Meng,~Y.-H.; Bai,~W.; Gao,~H.; Gong,~S.-J.; Wang,~J.-Q.; Duan,~C.-G.;
  Chu,~J.-H. Ferroelectric control of Rashba spin orbit coupling at the
  GeTe(111)/InP(111) interface. \emph{Nanoscale} \textbf{2017}, \emph{9},
  17957–17962\relax
\mciteBstWouldAddEndPuncttrue
\mciteSetBstMidEndSepPunct{\mcitedefaultmidpunct}
{\mcitedefaultendpunct}{\mcitedefaultseppunct}\relax
\EndOfBibitem
\bibitem[Jeong \latin{et~al.}(2021)Jeong, Lee, Lee, Wook, Kim, Lee, and
  Cho]{Jeong_2021}
Jeong,~K.; Lee,~H.; Lee,~C.; Wook,~L.~H.; Kim,~H.; Lee,~E.; Cho,~M.-H.
  Ferroelectric switching in GeTe through rotation of lone-pair electrons by
  Electric field-driven phase transition. \emph{Applied Materials Today}
  \textbf{2021}, \emph{24}, 101122\relax
\mciteBstWouldAddEndPuncttrue
\mciteSetBstMidEndSepPunct{\mcitedefaultmidpunct}
{\mcitedefaultendpunct}{\mcitedefaultseppunct}\relax
\EndOfBibitem
\bibitem[Nukala \latin{et~al.}(2017)Nukala, Ren, Agarwal, Berger, Liu, Johnson,
  and Agarwal]{Nukala_2017}
Nukala,~P.; Ren,~M.; Agarwal,~R.; Berger,~J.; Liu,~G.; Johnson,~A. T.~C.;
  Agarwal,~R. Inverting polar domains via electrical pulsing in metallic
  germanium telluride. \emph{Nature Communications} \textbf{2017}, \emph{8},
  15033\relax
\mciteBstWouldAddEndPuncttrue
\mciteSetBstMidEndSepPunct{\mcitedefaultmidpunct}
{\mcitedefaultendpunct}{\mcitedefaultseppunct}\relax
\EndOfBibitem
\bibitem[Croes \latin{et~al.}(2021)Croes, Cheynis, Zhang, Voulot, Dorkenoo,
  Cherifi-Hertel, Mocuta, Texier, Cornelius, Thomas, Richard, Müller,
  Curiotto, and Leroy]{Croes_2021}
Croes,~B.; Cheynis,~F.; Zhang,~Y.; Voulot,~C.; Dorkenoo,~K.~D.;
  Cherifi-Hertel,~S.; Mocuta,~C.; Texier,~M.; Cornelius,~T.; Thomas,~O.;
  Richard,~M.-I.; Müller,~P.; Curiotto,~S.; Leroy,~F. Ferroelectric
  nanodomains in epitaxial GeTe thin films. \emph{Physical Review Materials}
  \textbf{2021}, \emph{5}, 124415\relax
\mciteBstWouldAddEndPuncttrue
\mciteSetBstMidEndSepPunct{\mcitedefaultmidpunct}
{\mcitedefaultendpunct}{\mcitedefaultseppunct}\relax
\EndOfBibitem
\bibitem[Kresse and Hafner(1993)Kresse, and Hafner]{Kresse_1993}
Kresse,~G.; Hafner,~J. Ab initio molecular dynamics for liquid metals.
  \emph{Physical Review B} \textbf{1993}, \emph{47}, 558–561\relax
\mciteBstWouldAddEndPuncttrue
\mciteSetBstMidEndSepPunct{\mcitedefaultmidpunct}
{\mcitedefaultendpunct}{\mcitedefaultseppunct}\relax
\EndOfBibitem
\bibitem[Kresse and Furthmüller(1996)Kresse, and Furthmüller]{Kresse_1996a}
Kresse,~G.; Furthmüller,~J. Efficient iterative schemes for ab initio
  total-energy calculations using a plane-wave basis set. \emph{Physical Review
  B} \textbf{1996}, \emph{54}, 11169–11186\relax
\mciteBstWouldAddEndPuncttrue
\mciteSetBstMidEndSepPunct{\mcitedefaultmidpunct}
{\mcitedefaultendpunct}{\mcitedefaultseppunct}\relax
\EndOfBibitem
\bibitem[Kresse and Furthmüller(1996)Kresse, and Furthmüller]{Kresse_1996b}
Kresse,~G.; Furthmüller,~J. Efficiency of ab-initio total energy calculations
  for metals and semiconductors using a plane-wave basis set.
  \emph{Computational Materials Science} \textbf{1996}, \emph{6}, 15–50\relax
\mciteBstWouldAddEndPuncttrue
\mciteSetBstMidEndSepPunct{\mcitedefaultmidpunct}
{\mcitedefaultendpunct}{\mcitedefaultseppunct}\relax
\EndOfBibitem
\bibitem[Kresse and Joubert(1999)Kresse, and Joubert]{Kresse_1999}
Kresse,~G.; Joubert,~D. From ultrasoft pseudopotentials to the projector
  augmented-wave method. \emph{Physical Review B} \textbf{1999}, \emph{59},
  1758–1775\relax
\mciteBstWouldAddEndPuncttrue
\mciteSetBstMidEndSepPunct{\mcitedefaultmidpunct}
{\mcitedefaultendpunct}{\mcitedefaultseppunct}\relax
\EndOfBibitem
\bibitem[Perdew \latin{et~al.}(1996)Perdew, Burke, and Ernzerhof]{Perdew_1996}
Perdew,~J.~P.; Burke,~K.; Ernzerhof,~M. Generalized Gradient Approximation Made
  Simple. \emph{Physical Review Letters} \textbf{1996}, \emph{77},
  3865–3868\relax
\mciteBstWouldAddEndPuncttrue
\mciteSetBstMidEndSepPunct{\mcitedefaultmidpunct}
{\mcitedefaultendpunct}{\mcitedefaultseppunct}\relax
\EndOfBibitem
\bibitem[Grimme \latin{et~al.}(2010)Grimme, Antony, Ehrlich, and
  Krieg]{Grimme_2010}
Grimme,~S.; Antony,~J.; Ehrlich,~S.; Krieg,~H. A consistent and accurate ab
  initio parametrization of density functional dispersion correction (DFT-D)
  for the 94 elements H-Pu. \emph{The Journal of Chemical Physics}
  \textbf{2010}, \emph{132}, 154104\relax
\mciteBstWouldAddEndPuncttrue
\mciteSetBstMidEndSepPunct{\mcitedefaultmidpunct}
{\mcitedefaultendpunct}{\mcitedefaultseppunct}\relax
\EndOfBibitem
\bibitem[Henkelman \latin{et~al.}(2000)Henkelman, Uberuaga, and
  Jónsson]{Henkelman_2000}
Henkelman,~G.; Uberuaga,~B.~P.; Jónsson,~H. A climbing image nudged elastic
  band method for finding saddle points and minimum energy paths. \emph{The
  Journal of Chemical Physics} \textbf{2000}, \emph{113}, 9901–9904\relax
\mciteBstWouldAddEndPuncttrue
\mciteSetBstMidEndSepPunct{\mcitedefaultmidpunct}
{\mcitedefaultendpunct}{\mcitedefaultseppunct}\relax
\EndOfBibitem
\bibitem[Sheppard \latin{et~al.}(2012)Sheppard, Xiao, Chemelewski, Johnson, and
  Henkelman]{Sheppard_2012}
Sheppard,~D.; Xiao,~P.; Chemelewski,~W.; Johnson,~D.~D.; Henkelman,~G. A
  generalized solid-state nudged elastic band method. \emph{The Journal of
  Chemical Physics} \textbf{2012}, \emph{136}, 074103\relax
\mciteBstWouldAddEndPuncttrue
\mciteSetBstMidEndSepPunct{\mcitedefaultmidpunct}
{\mcitedefaultendpunct}{\mcitedefaultseppunct}\relax
\EndOfBibitem
\bibitem[Deringer \latin{et~al.}(2012)Deringer, Lumeij, and
  Dronskowski]{Deringer_2012}
Deringer,~V.~L.; Lumeij,~M.; Dronskowski,~R. Ab Initio Modeling of
  alpha-GeTe(111) Surfaces. \emph{The Journal of Physical Chemistry C}
  \textbf{2012}, \emph{116}, 15801–15811\relax
\mciteBstWouldAddEndPuncttrue
\mciteSetBstMidEndSepPunct{\mcitedefaultmidpunct}
{\mcitedefaultendpunct}{\mcitedefaultseppunct}\relax
\EndOfBibitem
\bibitem[Krempaský \latin{et~al.}(2016)Krempaský, Muff, Bisti, Fanciulli,
  Volfová, Weber, Pilet, Warnicke, Ebert, Braun, Bertran, Volobuev, Minár,
  Springholz, Dil, and Strocov]{Krempaský_2016}
Krempaský,~J.; Muff,~S.; Bisti,~F.; Fanciulli,~M.; Volfová,~H.; Weber,~A.~P.;
  Pilet,~N.; Warnicke,~P.; Ebert,~H.; Braun,~J.; Bertran,~F.; Volobuev,~V.~V.;
  Minár,~J.; Springholz,~G.; Dil,~J.~H.; Strocov,~V.~N. Entanglement and
  manipulation of the magnetic and spin–orbit order in multiferroic Rashba
  semiconductors. \emph{Nature Communications} \textbf{2016}, \emph{7},
  13071\relax
\mciteBstWouldAddEndPuncttrue
\mciteSetBstMidEndSepPunct{\mcitedefaultmidpunct}
{\mcitedefaultendpunct}{\mcitedefaultseppunct}\relax
\EndOfBibitem
\bibitem[Yang \latin{et~al.}(2021)Yang, Li, Li, Li, Sun, Liu, Bai, Li, Xie, Su,
  Gong, Zhang, He, and Cheng]{Yang_2021}
Yang,~X.; Li,~X.-M.; Li,~Y.; Li,~Y.; Sun,~R.; Liu,~J.-N.; Bai,~X.; Li,~N.;
  Xie,~Z.-K.; Su,~L.; Gong,~Z.-Z.; Zhang,~X.-Q.; He,~W.; Cheng,~Z.
  Three-Dimensional Limit of Bulk Rashba Effect in Ferroelectric Semiconductor
  GeTe. \emph{Nano Letters} \textbf{2021}, \emph{21}, 77–83\relax
\mciteBstWouldAddEndPuncttrue
\mciteSetBstMidEndSepPunct{\mcitedefaultmidpunct}
{\mcitedefaultendpunct}{\mcitedefaultseppunct}\relax
\EndOfBibitem
\bibitem[Gao \latin{et~al.}(2011)Gao, Nelson, Jokisaari, Baek, Bark, Zhang,
  Wang, Schlom, Eom, and Pan]{Gao_2011}
Gao,~P.; Nelson,~C.~T.; Jokisaari,~J.~R.; Baek,~S.-H.; Bark,~C.~W.; Zhang,~Y.;
  Wang,~E.; Schlom,~D.~G.; Eom,~C.-B.; Pan,~X. Revealing the role of defects in
  ferroelectric switching with atomic resolution. \emph{Nature Communications}
  \textbf{2011}, \emph{2}, 591\relax
\mciteBstWouldAddEndPuncttrue
\mciteSetBstMidEndSepPunct{\mcitedefaultmidpunct}
{\mcitedefaultendpunct}{\mcitedefaultseppunct}\relax
\EndOfBibitem
\bibitem[Picozzi(2014)]{Picozzi_2014}
Picozzi,~S. Ferroelectric Rashba semiconductors as a novel class of
  multifunctional materials. \emph{Frontiers in Physics} \textbf{2014},
  \emph{2}, 1\relax
\mciteBstWouldAddEndPuncttrue
\mciteSetBstMidEndSepPunct{\mcitedefaultmidpunct}
{\mcitedefaultendpunct}{\mcitedefaultseppunct}\relax
\EndOfBibitem
\bibitem[Liebmann \latin{et~al.}(2016)Liebmann, Rinaldi, Di~Sante, Kellner,
  Pauly, Wang, Boschker, Giussani, Bertoli, Cantoni, Baldrati, Asa, Vobornik,
  Panaccione, Marchenko, Sánchez-Barriga, Rader, Calarco, Picozzi, Bertacco,
  and Morgenstern]{Liebmann_2016}
Liebmann,~M.; Rinaldi,~C.; Di~Sante,~D.; Kellner,~J.; Pauly,~C.; Wang,~R.~N.;
  Boschker,~J.~E.; Giussani,~A.; Bertoli,~S.; Cantoni,~M.; Baldrati,~L.;
  Asa,~M.; Vobornik,~I.; Panaccione,~G.; Marchenko,~D.; Sánchez-Barriga,~J.;
  Rader,~O.; Calarco,~R.; Picozzi,~S.; Bertacco,~R.; Morgenstern,~M. Giant
  Rashba-Type Spin Splitting in Ferroelectric GeTe(111). \emph{Advanced
  Materials} \textbf{2016}, \emph{28}, 560–565\relax
\mciteBstWouldAddEndPuncttrue
\mciteSetBstMidEndSepPunct{\mcitedefaultmidpunct}
{\mcitedefaultendpunct}{\mcitedefaultseppunct}\relax
\EndOfBibitem
\end{mcitethebibliography}
\end{document}